\documentclass{ws-acs}

\newcommand{\dy}{{\rm dyn}}
\newcommand{\eq}{{\rm eq}}
\newcommand{\eps}{\varepsilon}
\newcommand{\sign}{\mathop{\rm sign}}
\def\draftnote{\today\quad\currenttime}
\def\catchline{}
\def\trimmarks{}
\begin{document}

\markboth{Peter F.\ Stadler, Anita Mehta, Jean-Marc Luck}
{Glassy states in a shaken sandbox}

\catchline

\title{Glassy states in a shaken sandbox}

\author{PETER F.\ STADLER}

\address{Institut f\"ur Theoretische Chemie und Molekulare
        Strukturbiologie,\\
        Universit\"at Wien,
        W{\"a}hringerstra{\ss}e 17, A-1090 Wien, Austria; and\\
        The Santa Fe Institute, 1399 Hyde Park Rd., Santa Fe NM 87501, USA
}

\author{ANITA MEHTA}

\address{S N Bose National Centre for Basic Sciences, Block JD Sector 3,
        Salt Lake, Calcutta 700098, India; and\\
        ICTP, Strada Costiera 11, I-34100 Trieste, Italy}

\author{JEAN-MARC LUCK}

\address{Service de Physique Th\'eorique (URA 2306 of CNRS), CEA Saclay,\\
91191 Gif-sur-Yvette cedex, France}

\maketitle

\pub{Received ()}{Revised ()}

\begin{abstract}
Our model of shaken sand, presented in earlier work,
has been extended to include a more realistic
`glassy' state, i.e., when the sandbox is shaken at very
low intensities of vibration. We revisit some of our earlier
results, and compare them with our new results on the revised model.
Our analysis of the glassy dynamics in our model shows
that a variety of ground states is obtained; these fall in two categories,
which we argue are representative of regular and irregular packings.
\end{abstract}

\section{Introduction}

The test of a good lattice model of a complex system is whether
it succeeds in capturing the essential physics of a real system
in its bid to reduce its technical complexity. Areas as diverse as traffic
flow~\cite{bml}, plate tectonics~\cite{bk,cl} and granular flow~\cite{Mehta:94}
are examples where lattice models have been used rather successfully
despite their apparent simplicity, to describe at least a few of the salient
features of some genuinely complex systems.

In this spirit, we present two versions of a model of shaken sand in the
following; both models exhibit behaviour that is representative of shaken sand
between the fluidised and the glassy regimes. While the second model includes
rather complex interactions between the `grains' in the frozen state, in
contrast to the first,
the latter is nevertheless surprisingly successful in replicating at least some
of the qualitative features associated with the glassy regime. One of the aims
of this contribution is then to identify some of the essential
features that are needed for a (discrete) minimal model of the system in
question.

The earlier model (`old model')~\cite{us:01} is
the generalisation of a cellular-automaton (CA)
model~\cite{Barker:00b,Mehta:94} of an avalanching sandpile.
This model shows {\em both} fast and slow dynamics in the
appropriate regimes: in particular, it reduces to an exactly solvable model
of noninteracting grains
in the frozen (`glassy') regime, and provides one with a toy model for
ageing in vibrated sand~\cite{letizia:00,Kurchan:00}.

Next, we present a more realistic version of the model
(`new model')~\cite{ustocome:01},
which is the topic of ongoing research.
While identical in the fluidised regime,
the latter model is less of a toy model for the glassy
state, in that the grains are no longer noninteracting, but are coupled
to each other based on their orientations. Our analysis shows that
a rich variety of ground states is obtained, which we analyse in terms
of a particular parameter that has an interpretation in terms of the
irregularity of the grains.

\section{The Model}

We consider a rectangular lattice of height $H$ and width $W$ with $N\le HW$
grains located at the lattice points, shaken with vibration intensity
$\Gamma$. Each grain is a rectangle with sides $1$ and $a\le1$,
respectively. Consider a grain $(i,j)$ in row $i$ and column $j$ whose
height at any given time is given by $h_{ij} = n_{ij-} + a n_{ij+}$, where
$n_{ij-}$ is the number of vertical grains and $n_{ij+}$ is the number of
horizontal grains below~$(i,j)$.

The dynamics of the model are described by the following rules:
\begin{itemize}
\item[(i)] If lattice sites $(i+1,j-1)$, $(i+1,j)$, or $(i+1,j+1)$ are
empty, grain $(i,j)$ moves there with a probability $\exp(-1/\Gamma)$, in
units such that the acceleration due to gravity, the mass of a grain, and
the height of a lattice cell all equal unity.
\item[(ii)] If the lattice site $(i-1,j)$ below the grain is empty, it will
fall down.
\item[(iii)] If lattice sites $(i-1,j\pm1)$ are empty, the grain at height
$h_{ij}$ will fall to either lower neighbour, provided the height
difference $h_{ij}-h_{i-1,j\pm1}\ge 2$.
\item[(iv)] The grain flips from horizontal to vertical with probability
$\exp(-m_{ij}(\Delta H +\Delta h)/\Gamma)$, where $m_{ij}$ is the mass of
the pile (consisting of grains of unit mass) above grain $(i,j)$. For a
rectangular grain, $\Delta H=1-a$ is the height difference
between the initial horizontal and the final vertical state
of the grain. Similarly, the {\em activation energy} for a flip reads
$\Delta h=b-1$, where $b=\sqrt{1+a^2}$ is the diagonal length of a grain.
\item[(v)] The grain flips from vertical to horizontal with probability
$\exp(\!-m_{ij}\Delta h/\Gamma)$.
\end{itemize}

\begin{figure}[t]
\centerline{\epsfig{file=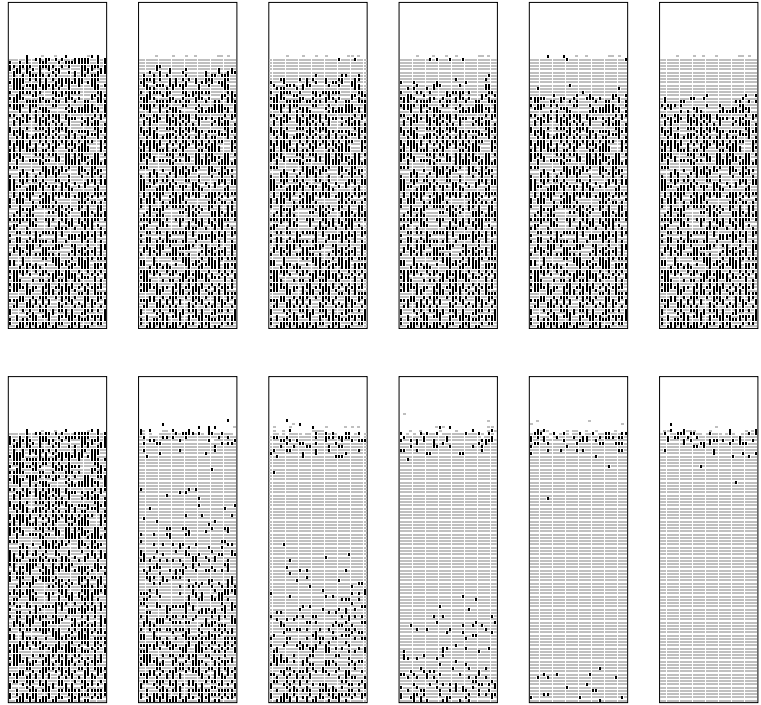,width=0.8\textwidth}}
\par\noindent
\caption{Snapshots of a sandbox with $W=30$, $H=100$, $n=2500$ grains with
$a=0.7$, $\Delta h=0.05$, and shaking intensities
$\Gamma=0.1$ (top row: glassy regime)
and $0.8$ (bottom row: fluidised regime),
for times $t=1$ (before tapping), $t=2$ (each
grain on average touched once by the MC simulation), $t=5$, $10$, $30$, and
$100$.}
\label{fig:snapshots}
\end{figure}

These rules yield a rich variety of dynamical behaviour, depicted in
Fig.~\ref{fig:snapshots}, which shows the time evolution of the sandbox under
the effect of different vibrational intensities. All the boxes are started
in the same initial state; the observed behaviours correspond to the glassy
regime (top row) and the fluidised regime (bottom row). In the
intermediate regime separating the two~\cite{us:01}, individual-particle
relaxations are the mechanism for each particle to find local stability.
In related work~\cite{bm2,bm1}, this was associated
with a threshold, called the {\em single particle relaxation threshold} (SPRT).

In line with recent investigations of
compaction~\cite{deGennes:2,Barker:92,Barker:93,deGennes:1,deGennes:3,
sam:1,sam:2,Mehta:91,NowakE:98,NowakE:97},
we have examined the behaviour
of the packing fraction of our model, as a function of the vibration
intensity $\Gamma$.
Let $N^-$ and $N^+$ be the numbers of vertical and
horizontal grains in the box.
The packing fraction $\phi$ is:
\begin{equation}
\phi=\frac{N^+-aN^-}{N^++aN^-},
\label{phidef}
\end{equation}
which we use as an order parameter reflective of the behaviour of the
compactivity~\cite{sam:1,sam:2}.
The vertical orientation of a grain thus
wastes space proportional to~$1-a$, relative to the horizontal one.

The advantage of our current model is that at least conceptually
it can be extended to non-rectangular grain shapes. In this sense,
we can and will consider~$a$, $\Delta H$, and $\Delta h$
as phenomenological parameters in the following.

\section{A Spin-Model for the Ordered Regime}

The frozen regime is characterised by an absence of holes within the
sandbox, and negligible surface roughness.
Here, the earlier model~\cite{us:01} reduces to an
exactly solvable model of $W$ independent columns of $H$ noninteracting
grain orientations $\sigma_n(t)=\pm1$, with $\sigma=+1$ denoting a
horizontal grain, and $\sigma=-1$ denoting a vertical grain.
The orientation of the grain at depth $n$, measured from the top of the system,
evolves according to a continuous-time Markov dynamics, with rates
\begin{equation}
\left\{\matrix{
w(-\to+)=\exp(-n\Delta h/\Gamma),\hfill\cr
w(+\to-)=\exp(-n(\Delta H+\Delta h)/\Gamma),\hfill}\right.
\end{equation}
as $m_{ij}=n=H+1-i$.
The parameters $\Delta H$ and $\Delta h$ correspond to
two characteristic lengths of the model,
\begin{equation}
\xi_\eq=\Gamma/\Delta H,\quad\xi_\dy=\Gamma/\Delta h.
\end{equation}
The equilibrium length $\xi_\eq$ is the typical depth below which
all grains are frozen into their horizontal ground-state orientation,
while the dynamical length $\xi_\dy$ is characteristic
of the divergence of the relaxation time with depth, $\tau\sim\exp(n/\xi_\dy)$.
As a consequence, any perturbation propagates only logarithmically slowly
down the system, over an ordering length $\Lambda(t)\approx\xi_\dy\ln t$.

The detailed analysis of this model in earlier work,
despite its `toy model' nature, was remarkably successful
in reproducing certain features of the glassy state. In this work,
we present an improved version where the spins are no longer noninteracting,
in the frozen regime. They are in fact coupled in a rather complex way;
these more realistic interactions lead to a rich variety of ground
states depending on the analogue of the aspect
ratio $a$, which we argue is representative
of the shape of the grains. We present this model in
the next section.

\section{A Generalisation}

The generalisation of our earlier model involves the insertion
of eq.~(2.1) into the transition rates of the system. We thus require
that, for a given value of $a$, the transitions are such that
the packing fraction of the system is locally minimised. We now allow
$a$ to take arbitrary values (while always remaining positive):
$1-a$ can then be visualised as the size of the `void' associated with the
`wrong' orientation of the grain. For convenience, we write
these rules in a more general form below.

Thus, the dynamical rules (iv) and (v) of the rectangular grain model may be
regarded as a special case of a more general model with transition rates
\begin{equation}
\left\{\matrix{
w(-\to+)=\exp(-(\lambda_{ij}+\eta_{ij})/\Gamma),\hfill\cr
w(+\to-)=\exp(-(\lambda_{ij}-\eta_{ij})/\Gamma),\hfill}\right.
\label{newrates}
\end{equation}
where
\begin{equation}
\eta_{ij}=Am_{ij+}+Bm_{ij-},\quad
\lambda_{ij}=Cm_{ij+}+Dm_{ij-},
\end{equation}
are, respectively, the ordering field and the activation energy
felt by grain~$(ij)$.
In these expressions,
$m_{ij\pm}$ is the number of horizontal and vertical grains above the
grain $(i,j)$, respectively. Note that grains are only counted from $(i,j)$
to the first void above level $j$. Thus $m_{ij}=m_{ij+}+m_{ij-}$.

The earlier model is recovered by setting
\begin{equation}
A=B=\Delta H/2,\quad C=D=\Delta H/2+\Delta h.
\end{equation}
In the general situation where the equalities $A=B$, $C=D$ are not obeyed,
the rates~(\ref{newrates}) depend on the orientations of all the grains
above the grain under consideration.
Our new model is therefore a fully directed model of interacting grains,
where causality acts both in time and in space,
as the orientation of a given grain only influences the grains below it
and at later times.

The key parameter which governs the statics and dynamics of the model
at low shaking intensity turns out to be the dimensionless ratio $\eps=A/B$.
Consider the ordered regime, where there are no holes,
in the zero-temperature limit ($\Gamma\to0$).
In this regime, the steady-state values of the grain orientations
are given by the deterministic, recursive equation
\begin{equation}
\sigma_n=\sign(\eps n_+(n)+n_-(n)),
\label{deter}
\end{equation}
where $n_\pm(n)$ is the number of horizontal and vertical grains
at depths $0,\dots,n-1$, so that $n_+(n)+n_-(n)=n=1,2,\dots$
We assume $\sigma_0=+1$.

As long as $\eps\ge0$,~(\ref{deter}) leads to the trivial ground state
where all the grains are horizontal ($\sigma_n=+1$),
generalising thus the case of the earlier model $(\eps=1)$.

In the frustrated regime ($\eps<0$),
ground states have a richer structure.
They contain non-trivial fractions of horizontal and vertical grains,
\begin{equation}
f_+=1/(1-\eps),\quad f_-=-\eps/(1-\eps).
\end{equation}
The grain orientations are distributed in a way which depends,
rather unexpectedly, on whether $\eps$ is rational or not.

\begin{itemize}
\item Rational case:
if $\eps=-p/q$ is a negative rational number (in irreducible form),
(\ref{deter}) determines the grain orientation $\sigma_n$
whenever the depth $n$ is not a multiple of $r=p+q$.
The orientations of the latter grains are left free.
We thus obtain an extensively degenerate set of ground states,
each of them being a random sequence of two types of
$r$-mers, i.e., clusters of $r$ grains.
The associated configurational entropy per grain reads $\Sigma=(\ln 2)/r$.
The simplest example is $\eps=-1$, hence $r=2$ and $f_-=1/2$,
where the clusters are $+-$ and $-+$, so that the ground states
are all the dimerised grain configurations.
Two examples consist of trimers ($r=3$), namely $\eps=-1/2$,
with $f_-=1/3$ and clusters $+-+$ and $-++$,
and $\eps=-2$, with $f_-=2/3$ and clusters $+--$ and $-+-$.

\item Irrational case:
if $\eps$ is a negative irrational number,
(\ref{deter}) determines all the grain orientations,
so that the model admits a unique, non-degenerate ground state,
where the grain orientations are distributed in a quasiperiodic fashion.
The rule~(\ref{deter}) is indeed equivalent to the
cut-and-project algorithm used to build quasiperiodic binary chains,
which are one-dimensional analogues of perfect
quasicrystals~\cite{de:81,du:85,du:86,el:85,kkl2:85,kkl1:85,lgjj:93}.
For instance, for $-\eps=(\sqrt5-1)/2\approx0.618033$
(the inverse golden mean),
the unique ground state of the model is given by
the well-known Fibonacci sequence:
\begin{equation*}
+-+-++-+-++-++-+-++-+-++-++-+-+\cdots
\end{equation*}
\end{itemize}

\section{Numerical Results}

In Fig.~\ref{fig:phi} we show the behaviour of the
packing fraction as a function of $\Gamma$, the shaking intensity, for both
models described above for a square box of side $120$, containing $10000$
grains of unit mass, each of which has an aspect ratio of $0.7$.

We note that the overall behaviour of the two models is rather similar,
although the complexity of the second makes it far more prone to
fluctuations even after the steady state has been reached. In each case,
we observe
\begin{itemize}
\item a fluidised region: for $\Gamma\gg 1$, we observe an initial
increase (caused by a {\it non-equilibrium} and transient `ordering' of
grains in the boundary layer) of the packing fraction that quickly relaxes
to the equilibrium values $\phi_{\infty}$ in each case. This over-shooting
effect in Fig.~\ref{fig:phi} increases with $\Gamma$, since grains ever
deeper in the sandbox can now overcome their activation energy to relax to
the horizontal.
\item an intermediate region (for $\Gamma\approx 1$), where the packing
fraction remains approximately constant in the bulk, while the surface
equilibrates via the fast dynamics of {\em single-particle relaxation}. The
specific $\phi_{\infty}$ at which this occurs, is the {\em
single-particle relaxation threshold density} observed in
Ref.~\cite{bm1}; non-equilibrium, non-ergodic, fast dynamics allows
single particles locally to find their equilibrium configurations at this
density. Analogous effects have been observed in recent experiments on
colloids~\cite{science:00}, where the correlated dynamics of {\em fast}
particles was seen to be responsible for most relaxational behaviour before
the onset of the glass transition.
\begin{figure}[t]
\par\noindent
\centerline{\epsfig{file=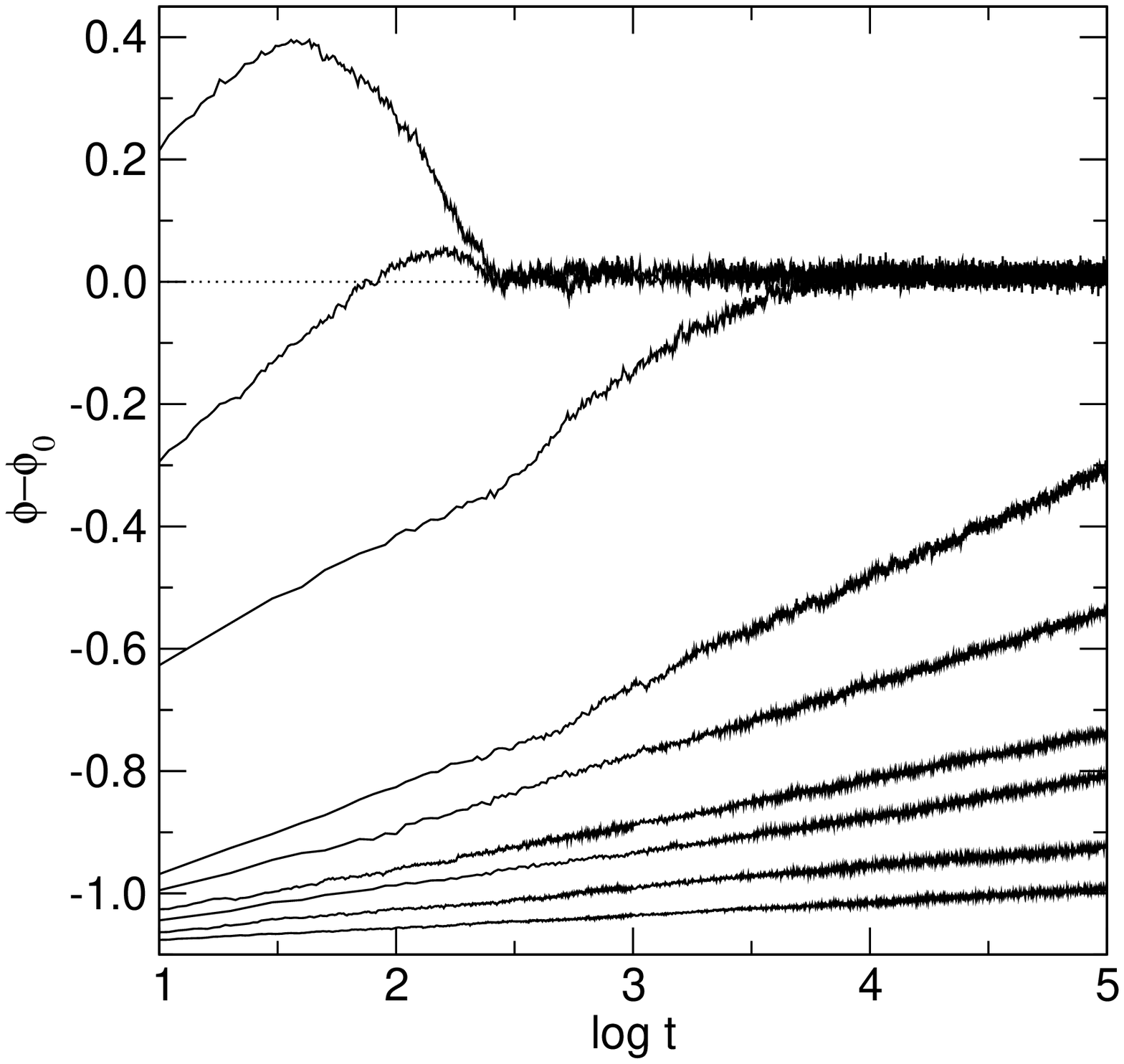,width=0.48\textwidth,clip=}
\psfig{file=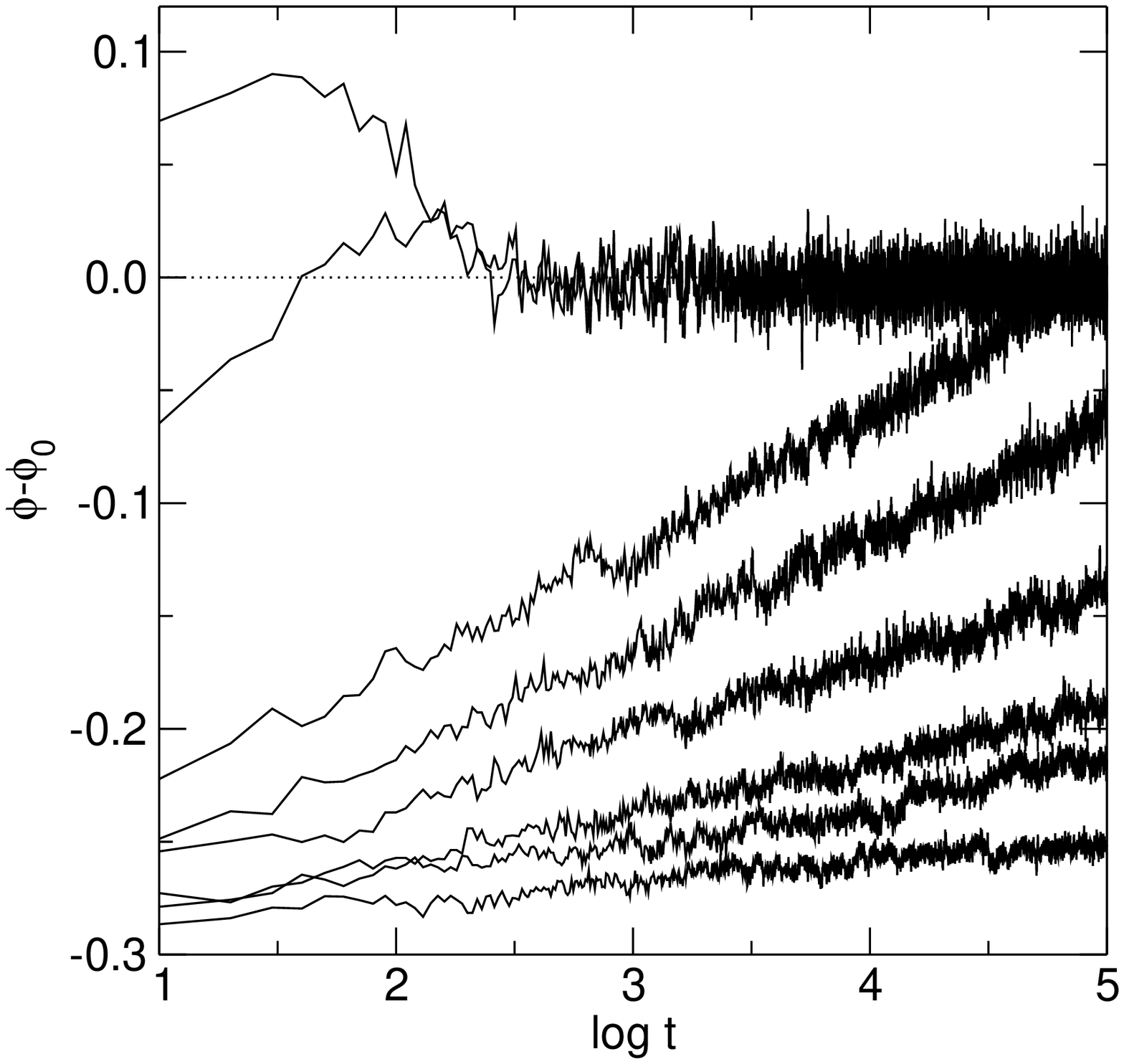,width=0.48\textwidth,clip=}}
\par\noindent
\caption{Behaviour of packing fraction for both models described in text,
with an aspect ratio of $0.7$ in each case. For the old model (left),
$A=B=0.35$, $C=D=0.4$, while for the new model (right)
$A=-0.2$, $B=C=D=0.4$. The
shaking intensities are, for bottom to top, $\Gamma=0.3$ (old model only),
$0.5$, $0.8$, $1$, $1.5$, $2$, $5$, and $10$.}
\label{fig:phi}
\end{figure}
\item a frozen region (for $\Gamma\ll 1$), where the slow dynamics of the
system results in a {\em logarithmic growth} of packing fraction
with time:
\begin{equation}
\phi-\phi_{\infty} = b(\Gamma)\ln t + a,
\end{equation}
where $b(\Gamma)$ increases with $\Gamma$, in good agreement with
experiment~\cite{NowakE:98,NowakE:97}.
The slow dynamics has been identified~\cite{bm1}
with a cascade process, where the free volume released
by the relaxation of one or more grains allows for the ongoing relaxation
of other grains in an extended neighbourhood. It includes the phenomenon
of {\it bridge collapse}, which, for low vibration intensities, has been
seen to be a major mechanism of compaction~\cite{Barker:92,Barker:93,Mehta:91}.
As $\Gamma$
decreases, the corresponding $\phi_{\infty}$ increases asymptotically
towards the jamming limit $\phi_{\mathrm{jam}}$, identified with a {\it
dynamical phase transition} in related work~\cite{johannes:this_volume}.
\end{itemize}

While we have presented in earlier work~\cite{us:01} a full analysis of
two-time correlation functions for the old model, work is currently
in progress to investigate this in the rather more complex new model.

We finally investigate the analogue of `annealed cooling', where $\Gamma$ is
increased and decreased cyclically, and the response of the packing
fraction observed~\cite{NowakE:98,NowakE:97}. The results obtained
here are similar to those~\cite{unpublished:00} seen using more realistic
models of shaken spheres, but the simplicity of the present lattice-based
models allows for a greater transparency.
\begin{figure}[ht]
\par\noindent
\centerline{\epsfig{file=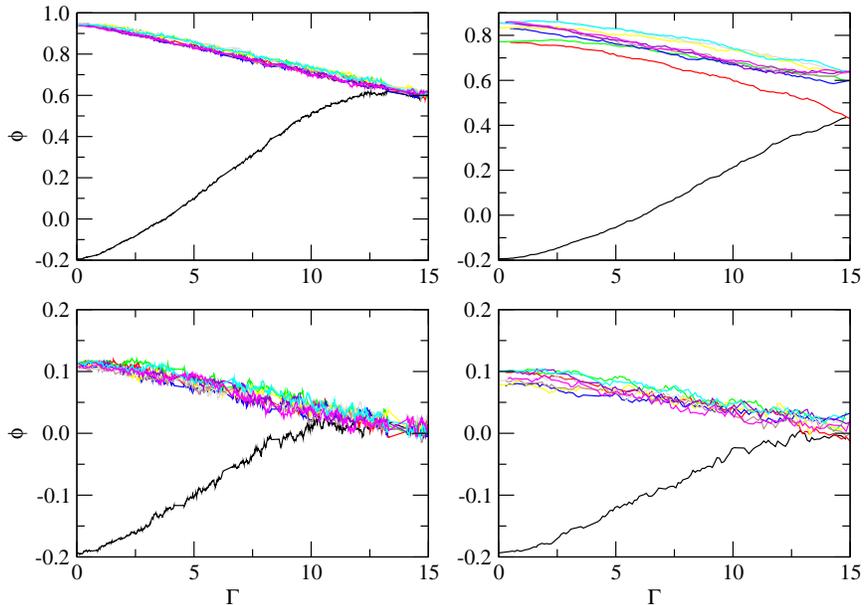,width=0.9\textwidth,clip=}}
\par\noindent
\caption{Hysteresis curves.
Top row: old model.
Bottom row: new model.
Left: $\delta\Gamma=0.1$, $t_{\mathrm{tap}}= 2000$ time units.
Right: $\delta\Gamma=0.001$, $t_{\mathrm{tap}}= 10^5$ time units.
Note the approach
of the irreversibility point $\Gamma^{*}$ to the `shoulder'
$\Gamma_{\mathrm{jam}}$, as the ramp rate
$\delta\Gamma/t_{\mathrm{tap}}$ is lowered. The packing fraction
tends to its close-packing limit in the limit of low intensities for the old model,
while it asymptotes towards the jamming limit for the new model (see text).}
\label{fig:fig3}
\end{figure}

Starting with the sand in a fluidised state, as in the
experiment~\cite{NowakE:98,NowakE:97},
we submit the sandbox to taps at a given
intensity $\Gamma$ for a time $t_{\mathrm{tap}}$ and increase the
intensity in steps of $\delta\Gamma$; at a certain point, the cycle is
reversed, to go from higher to lower intensities. The entire process is
then iterated twice. Fig.~\ref{fig:fig3} shows the resulting behaviour of
the volume fraction $\phi$ as a function of $\Gamma$, where an
`irreversible' branch and a `reversible' branch of the compaction curve are
seen, which meet at the `irreversibility point'
$\Gamma^{*}$~\cite{NowakE:98,NowakE:97}.
The left- and right-hand side of Fig.~\ref{fig:fig3} correspond
respectively to high and low values of the `ramp rate'
$\delta\Gamma/t_{\mathrm{tap}}$~\cite{NowakE:98,NowakE:97},
while the upper and lower
panels correspond respectively to the old and new versions of our model.
As the ramp rate is lowered, we note that:
\begin{itemize}
\item the width of the hysteresis loop in the so-called reversible branch
decreases, in both cases. The `reversible' branch is thus not reversible
at all; more realistic simulations
of shaken spheres~\cite{Barker:92,Barker:93,Mehta:91} confirm the first-order,
irreversible nature of the transition, which allows the density to attain
values that are substantially higher than random close packing, and quite
close to the crystalline limit~\cite{Mehta:00}. Precisely such a transition
has also recently been observed experimentally in the compaction of
rods~\cite{villaruel:00}.
\item In both panels, the `irreversibility point' $\Gamma^{*}$ approaches
$\Gamma_{\mathrm{jam}}$ (the shaking intensity at which the jamming limit
$\phi_{\mathrm{jam}}$ is approached), in agreement with results on other
discrete models~\cite{coniglio:00}.
However, in the upper panel, the packing fraction at low intensities
tends towards close-packing (the so-called dynamical transition referred
to in \cite{bm1});
 this is  at odds with
the results of real experiment, which models with a greater degree of
complexity~\cite{bm2} have been able to replicate. In the lower
panel, which corresponds to our new model,
we see tentative indications of
an improvement in this respect vis-a-vis the old model; the packing
fraction here asymptotes
 towards the jamming limit, rather than
rising indefinitely towards the close-packing limit
\cite{bm2,NowakE:98}.
\end{itemize}

\section{Discussion}

We have presented in the above two models of shaken sandboxes; while
their design was such that they would show identical behaviour
in regimes where there were a finite density of voids, the modelling
of the densely packed regime was completely distinct in each case.
In the first case, the model reduced to a model of noninteracting spins,
while in the second, the insistence that all allowed transitions
minimised a suitably defined local packing fraction led in fact
to an intricate coupling between the grains. This physically motivated
interaction was extremely nonlocal as well as directional.
In this way
we were able to generate a model that, despite being one-dimensional,
has an extremely complex ground-state structure,
depending on the {\em regularity} of the grain shape.
Our present investigations, to be published elsewhere~\cite{ustocome:01},
concern the effect
of zero-temperature and finite-temperature tapping of this system;
our preliminary studies indicate that for regularly shaped
grains, strong metastability in the achievable ground states is observed.
For irregularly shaped grains, as in reality, a far better packing
is achievable, since orientations of irregularly shaped grains
are much better able to fill space~\cite{torqchaikin}.

It is however a rather salutary exercise to see
that despite the relative sophistication of the new model
in its inclusion of non-trivial interactions, most of the qualitative behaviour
of the packing fraction as a function of steady as well as annealed
tapping, remains
unchanged. We expect that quantitative features such as two-time correlation
functions will be far more non-trivial in the second model
than the first, although we expect their overall features to be
rather similar. It is tempting
to speculate that the directionality due to gravity (which leads to strongly
non-Hamiltonian behaviour,
since grain couplings are propagating down the pile)
which unites both first and second
models might well be the most important ingredient that is needed to describe
such lattice-based models of shaken sand.

\bibliography{sandbox}
\bibliographystyle{acs}
\end{document}